\documentclass[12pt,epsf]{article}
\usepackage{graphicx, amsmath}

\begin{document}

\sloppy
\begin{flushright}{SIT-HEP/TM-30}
\end{flushright}
\vskip 1.5 truecm
\centerline{\large{\bf Elliptic Inflation: }}
\centerline{\large{\bf Generating the curvature perturbation without 
slow-roll }}
\vskip .75 truecm
\centerline{\bf Tomohiro Matsuda
\footnote{matsuda@sit.ac.jp}}
\vskip .4 truecm
\centerline {\it Laboratory of Physics, Saitama Institute of
 Technology,}
\centerline {\it Fusaiji, Okabe-machi, Saitama 369-0293, 
Japan}
\vskip 1. truecm
\makeatletter
\@addtoreset{equation}{section}
\def\theequation{\thesection.\arabic{equation}}
\makeatother
\vskip 1. truecm

\begin{abstract}
\hspace*{\parindent}
There are many inflationary models in which inflaton field does
 not satisfy the slow-roll condition.
However, in such models, it is always difficult to generate the curvature
 perturbation during inflation.
Thus, to generate the curvature perturbation, one must introduce another
component to the theory.
To cite a case, curvatons may generate dominant part of the curvature
perturbation after inflation.
However, we have a question whether it is unrealistic to
consider the generation of the curvature perturbation during 
inflation without slow-roll.
Assuming multi-field inflation, we encounter the generation of
the curvature perturbation during inflation without slow-roll.
The potential along equipotential surface is flat by definition
and thus we do not have to worry about symmetry.
We also discuss about KKLT models, in which corrections lifting
the inflationary direction may not become a serious problem if
there is a symmetry enhancement at the tip (not at the moving brane)
of the inflationary throat.
\end{abstract}

\newpage
\section{Introduction}
Among many benefits from the inflationary expansion that takes
place in the early
Universe, an important prediction of inflation would be the generation
of a spectrum of the primordial perturbations.
Such perturbations naturally arise from the zero point vacuum
fluctuations in quantum fields, which are stretched during inflation to
cover very large scales in our present Universe.
In the standard scenario of the inflationary Universe, the observed
density perturbation is produced by a light inflaton field 
that rolls slowly down its potential.
At the end of inflation, the inflaton field oscillates about the minimum
of its potential and decays to reheat the Universe.
Adiabatic density perturbation is generated because the scale-invariant 
fluctuations
of the light inflaton field are different in different patches.
Generically, slow-roll models of inflation predict an almost
scale-invariant and Gaussian distribution of primordial density
perturbations.

We are not going back to historical developments, however 
it is easy to understand that the idea of hybrid inflation 
may provide us with a key idea to construct successful inflationary
models\cite{EU-book}. 
In hybrid models, the end of inflationary expansion is a
second-order phase 
transition triggered by a trapped field (waterfall field).
D-term inflation is an important application of this idea, which is
found in the paradigm of supersymmetric particle cosmology.
Furthermore, an important variant of D-term inflation is found in the
paradigm of brane cosmology, which is called brane
inflation\cite{brane-inflation0, angled-inflation,
matsuda_braneinflation}. 
From phenomenological viewpoints, brane models are sometimes categorized
as models with large or intermediate extra dimensions (and vice
versa)\cite{Extra_1}.
The idea of large extra dimensions is important for higher-dimensional
models, because it may solve or weaken the hierarchy problem. 
In models with large extra dimensions, fields in the standard model(SM)
are localized on a wall-like structure (maybe it is a brane), 
while the graviton propagates in the bulk.
The discrepancy of the volume factor between gauge fields and
gravitational fields explains the large
hierarchy between gravity and gauge interactions.

Of course, it is an important challenge to find signatures of branes in 
cosmological observations.
Historically, it has been discussed that studying
the formation and the evolution of 
cosmological defects would provide us with important information about
branes\cite{brane-defects, matsuda-defects}.\footnote{Supergravity
provides a natural mechanism for removing cosmological domain
walls\cite{matsuda-wall}.} 
Inflation models with low fundamental scale are   
 discussed in ref.\cite{low_inflation}.
Scenarios of baryogenesis in such low-scale models are discussed in
ref.\cite{low_baryo, Defect-baryo-largeextra,  
Defect-baryo-4D}, where defects play distinguishable roles.
The curvatons\cite{curvaton_1, curvaton_2, curvaton_3} will play
significant roles in 
these low-scale models\cite{curvaton_liberate, topologicalcurvaton}. 
Moreover, in ref.\cite{topologicalcurvaton}, it has been discussed
that 
topological defects can play the role of the curvatons.
Thus, defects in brane models such as monopoles, strings, domain walls and
Q-balls are important\cite{matsuda_necklace, BraneQball,
matsuda_monopoles_and_walls, incidental, 
matsuda_angleddefect, tit-new}.
It might be important to explain why 
defects other than strings can be produced in brane inflationary models,
since (historically) it has been discussed by many
authors that only
strings are produced in brane inflationary
models\cite{previous-onlystrings}.
It is not hard to understand that 
this conjecture is not so generic as it has been anticipated,
as one can see from ref.\cite{matsuda_JGRG}.
Therefore, in order to find signatures of branes, it is important to
consider other types of defects as well as cosmic strings.
Our scenario of elliptic inflation is important, in a sense that it may
provide us with another clue as to how one can find signature of branes
from cosmological observations, as we will discuss in Sect.3.5.

When it appeared, D-term inflation was believed to circumvent the
well-known eta-problem of supergravity models of inflation based on
F-terms.
Thus brane inflation, which is in a sense a variant of
D-term inflation, was believed to circumvent the eta-problem.
However, a similar problem arose later again in brane paradigm.
The problem was that the very mechanism that lifts the moduli potential
was found to lift inflaton candidates and violate slow-roll
conditions\cite{KKLT-eta}. 
Directions that are protected by exact global symmetry may be expected to
remain flat, however it is still very difficult to find actual symmetry 
that can survive after moduli stabilization and can be used to construct
a successful inflationary scenario.\footnote{There are
interesting approaches in ref.\cite{DBI-inflation} and \cite{Sinha}. }
From the perspective we discussed above, we believe that 
finding new mechanism for generating the curvature perturbation without
slow-roll inflaton is quite important.
In general, the superhorizon spectrum of perturbations is thought to be
generated by the amplification of the quantum fluctuations of a light
inflaton field, whose mass is much smaller than the Hubble constant during
inflation. 
This is because the quantum fluctuations of the field can reach and exit
the horizon only if its Compton wavelength is larger than the horizon
during inflation.
However, this happens only if the inflaton is effectively massless,
i.e. only if $m_{I} \ll H_I$.
The above condition seems to conflict with our aim in this paper.
Luckily, we know alternatives for the slow-roll inflaton.
The curvatons and their variants can generate the curvature
perturbation after inflation\cite{curvaton_1, curvaton_2},
even if there is no slow-roll during inflation.
However, this mechanism still requires flat direction that is supposed
to have non-trivial properties.
Thus, we think it is still important to find models for generating
the curvature perturbation without using neither slow-roll inflaton 
nor the curvatons. 
The situations that we will consider in this paper
are both general and useful.
Our mechanism can be utilized in many kinds of phenomenological models
in which slow-roll condition is inevitably violated.
As we are considering inflation without slow-roll, the e-foldings of
our model may be short, which will require compensation by another
period of inflationary expansion.

\section{Elliptic Inflation} 
Consider two inflaton fields $(\phi_1, \phi_2)$ with a hybrid type of
potential of the form\footnote{In this section we outline a
comprehensive strategy for obtaining 
inflationary expansion without slow-roll.
One may be disappointed to see what everyone has already seen.
However, this is not the point that we want to discuss in this
paper. 
What we want to address in this paper is not the way how one can obtain
inflationary expansion without slow-roll, but how one can obtain the
curvature perturbation without neither slow-roll inflaton nor the 
curvatons.
We will consider the latter issue in Sect.3.}
\begin{equation}
\label{pot-full}
V(\phi_1, \phi_2, \sigma)=
\frac{1}{2}m^2\left(\phi_1^2 + A \phi_2^2\right)
+\frac{\lambda_1}{2}\left(\phi_1^2 + B \phi_2^2\right) \sigma^2
+\frac{\lambda_2}{4}\left(\sigma^2-M^2\right),
\end{equation}
where $\phi_i$ $(i=1,2)$ and $\sigma$ are taken to be real scalar fields.
For later convenience, we introduce a field $\phi_r$
\begin{equation}
\phi_r^2 \equiv \phi_1^2 +A \phi_2^2
\end{equation}
and mass 
\begin{equation}
m_\sigma^2 \equiv \lambda_2 M^2.
\end{equation}
The above potential has global minima at 
\begin{equation}
(\phi, |\sigma|)=(0,M),
\end{equation}
and an unstable saddle point at 
\begin{equation}
(\phi, \sigma)=(0,0).
\end{equation}
One may be disappointed to see the above potential, since it
looks quite similar to the one that has been used in usual hybrid
inflation.\footnote{Hybrid inflation without
slow-roll is already discussed in ref.\cite{Dimopoulos-Fast-Osc} for
single-field inflation.}
However, we never assume slow-roll condition for the inflation field,
except for ``built-in models'' that we will consider in Sect.3.4

Since the effective mass squared of the waterfall field $\sigma$
is given by $m_\sigma^2(\phi)=\lambda_1 \left(\phi_1^2 + B \phi_2^2
\right) 
-m_\sigma^2$,
the waterfall field $\sigma$ remains at $\sigma=0$ as far as
the following condition is satisfied;
\begin{equation}
\label{range}
\phi_1^2 + B \phi_2^2 >\phi_c^2\equiv m^2_\sigma/\lambda_1
\end{equation}
Suppose that initially the system lies in the above
region (\ref{range})
and $\sigma$ is at the quasi-stable point $\sigma =0$.
In the case where the false vacuum density is dominated by a scalar
potential 
\begin{equation}
\label{pot1}
V_{I} = \frac{\lambda_2}{4} M^4 + \frac{1}{2} m^2\phi_r^2,
\end{equation}
and if $\frac{\lambda_2}{4} M^4 > \frac{1}{2} m^2\phi_r^2$,
there would be a period of inflationary expansion even if there is no
slow-roll inflaton field.
During this period, according to the Friedman equation,
we have $H_I \simeq V_{I}^{1/2}/M_p$.
Since we are not expecting slow-roll inflation, the Hubble constant
during inflation can be much smaller than the masses of the
fields $\phi_i, i=1,2$.\footnote{We consider two fields,
 $\phi_1$ and $\phi_2$.
We know both fields may play the role of inflaton, however the actual
e-foldings is almost determined by either field.
Thus, we use the word ``inflaton'' for the field that determines the
total e-foldings.
In this paper, we use the field $\phi_1$ as the inflaton.}
During the period of inflation, the Klein-Gordon equation for $\phi_i$
is
\begin{equation}
\label{Klein-Gordon}
\ddot{\phi}_{i}+3H\dot{\phi}_i +\frac{\partial V}{\partial \phi_i}=0,
\end{equation}
where dots denote the derivatives with respect to the cosmic time $t$.
The above equation has solution for the background dynamics that can be
computed.
The background field dynamics is given by
\begin{equation}
\label{KG-eq}
\phi_i(t) \propto e^{\alpha_i t},
\end{equation}
where
\begin{equation}
\label{KG-alpha}
\alpha_i = \frac{3}{2}H_I 
\left[ 1\pm \left\{1-\frac{4m_i^2}{9H_I^2}\right\}^{1/2}\right].
\end{equation}
Here the masses of the fields are denoted by $m_i$.

\subsection{Fast-roll}
\begin{figure}[h]
 \begin{center}
\begin{picture}(400,220)(0,0)
\resizebox{10cm}{!}{\includegraphics{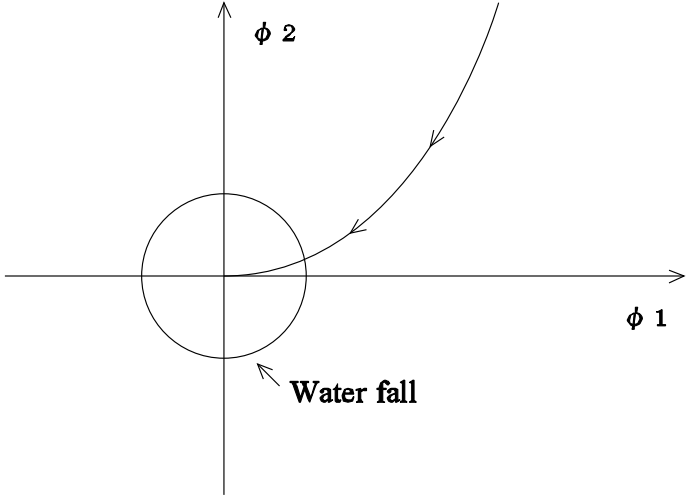}} 
\end{picture}
\caption{A trajectory $(\phi_1(t), \phi_2(t))$
during inflationary expansion.
We took a specific condition $m_i < \frac{3}{2}H_I$ (fast-roll) for
$i=1,2$. 
In the case where $\phi_2$ becomes sufficiently small before the end of
  inflation, the total e-foldings are determined by the evolution of
  $\phi_1$.
We know in such cases one tends to think that the secondary field $\phi_2$
plays no role in generating the curvature perturbation.
However, this is not the case, as we will explain below.} 
\label{Fig:fast-fast}
 \end{center}
\end{figure}

The idea of fast-roll inflation is first advocated by Linde in
ref.\cite{fast-roll-original}.
Fast-roll inflation occurs during a period of inflation 
when the mass of the inflaton field is 
comparable to the Hubble parameter.
In our case, as suggested by eq.(\ref{KG-alpha}), fast-roll occurs
when 
\begin{equation}
m_i \le \frac{3}{2} H_I.
\end{equation}
There are two solutions to the Klein-Gordon equation (\ref{KG-eq}),
both of which are exponentially decreasing with time.
Here we use the one with the negative sign, since the one with the
opposite sign corresponds to the fast-decaying mode that rapidly
disappears.
Therefore, the solution that survives is 
\begin{equation}
\phi_i = \phi_{i0}e^{-F_i Nt},
\end{equation}
where $N(t)$ is the number of the e-folds elapsed during inflation, and
$F_i$ is given by
\begin{equation}
F_i \equiv \frac{3}{2} 
\left[ 1- \left\{1-\frac{4m_i^2}{9H_I^2}\right\}^{1/2}\right].
\end{equation}
Assuming that (1) $\phi_1$ comes later than $\phi_2$, and 
(2) $\phi_1$ triggers waterfall
stage when $\phi_2$ is already in the region $|\phi_2| \ll
\phi_c/\sqrt{B}$,  
the total number of
e-folds elapsed during fast-roll inflation is given by
\begin{equation}
N_F =F_{1}^{-1}ln\left(\frac{\phi_{1,0}}{\phi_{1e}}\right)
\end{equation}
where $\phi_{1e}^2 = \phi_c^2 -B\phi_{2e} \simeq \phi_c^2$ denotes 
the end-point of fast-roll inflation.
See also Fig.\ref{Fig:fast-fast}.

\subsection{Locked oscillation}
Locked inflation was advocated in ref.\cite{locked1} using
a potential of the form eq.(\ref{pot-full}).
In the original model of locked inflation, an inflaton field 
oscillates on top of the false vacuum $\sigma =0$ during inflationary
expansion.
As is seen from (\ref{KG-alpha}), the boundary between fast-roll
inflation and locked inflation is at $m_i=\frac{3}{2} H_I$.
In the case where $m_i>\frac{3}{2}H_I$, one obtains oscillating inflaton
field that locks $\sigma$ at the origin.
Then, the Klein-Gordon equation is solved by an equation
\begin{equation}
\phi_i = \overline{\phi}_i(t) \cos(\beta_i t),
\end{equation}
where
\begin{equation}
\beta_i = H_I\sqrt{\frac{m_i^2}{H_I^2}-\frac{9}{4}}.
\end{equation}
The amplitude of the oscillation decreases with time, which becomes
\begin{equation}
\label{barphi}
\overline{\phi_i}(t)=\phi_{i,0}e^{-\frac{3}{2}N(t)}.
\end{equation}
The total number of e-foldings is given by
\begin{equation}
N_L(t) =\frac{2}{3}ln\left(\frac{\phi_{10}}{\phi_{1e}}\right),
\end{equation}
where we neglected effects from parametric resonance, which may or may
not be significant depending on the model we will consider.

\begin{figure}[h]
 \begin{center}
\begin{picture}(400,250)(0,0)
\resizebox{10cm}{!}{\includegraphics{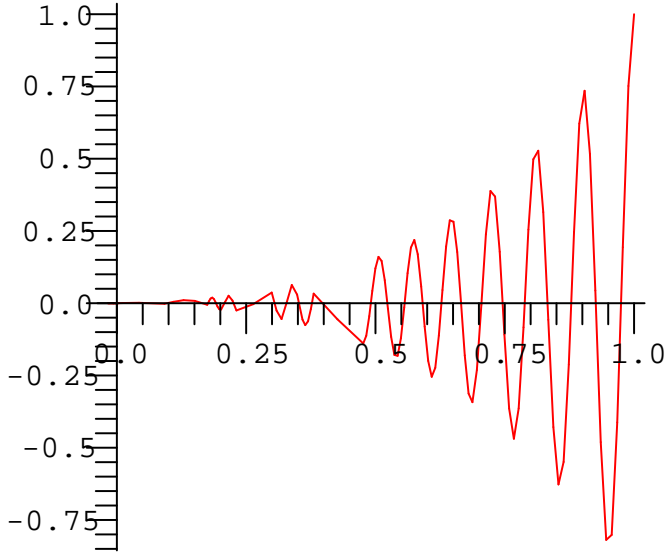}} 
\end{picture}
\caption{Artist's view of a trajectory $(\phi_1(t), \phi_2(t))$
during inflationary expansion.
We took a specific condition $m < \frac{3}{2}H_I$(fast-roll) for 
$\phi_1$
 and $\sqrt{A} m > \frac{3}{2}H_I$(locked) for $\phi_2$.
In this case, parametric resonance is suppressed even if the 
oscillation of the field $\phi_2$ is efficient at the beginning, 
since $\phi_2$ cannot hit the ``core''. }
\label{Fig:osc_fast}
 \end{center}
\end{figure}

It may be important to note that there is a significant discrepancy
between the usual oscillation in locked models and the one considered in
our model. 
In our case, oscillating field does
not necessarily hit the core 
region (i.e. $\sqrt{\phi_1^2 + B \phi_2^2} < \phi_c$) 
during its oscillation.
We show in Fig.\ref{Fig:osc_fast} what happens if the 
oscillation occurs during fast-roll inflation.

\subsection{Fast-roll after slow-roll}
\begin{figure}[h]
 \begin{center}
\begin{picture}(400,250)(0,0)
\resizebox{10cm}{!}{\includegraphics{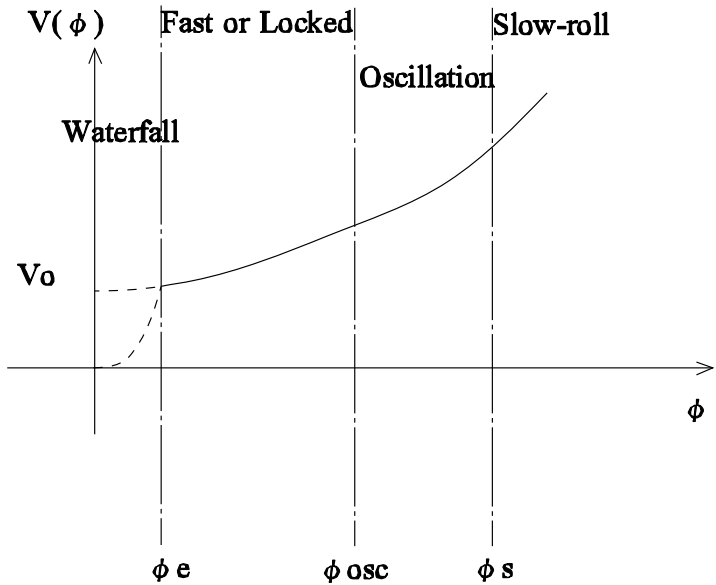}} 
\end{picture}
\caption{$\phi$ rolls slowly at a distance $\phi>\phi_s\sim M_p$, 
where mass term dominates the vacuum energy.
Mass term still dominates the vacuum energy if $\phi$ is larger than
  $\phi_{osc}$, where $\phi_{osc}$ is given by $\phi_{osc}\equiv \sqrt{2V_0}/m$. 
During the intermediate stage $\phi_{osc} <\phi <\phi_s$ where mass term
dominates the vacuum energy while slow-roll condition is violated,
$\phi$ oscillates like curvatons.
Then, fast-roll (or locked) inflation starts at $\phi=\phi_{osc}$.}
\label{Fig:slow-osc}
 \end{center}
\end{figure}

One may consider a case where the vacuum energy (\ref{pot1})
 is dominated by  a mass term of the field $\phi_1$.
This may happen in the outer region where $\phi_1$ is
 large.
See also Fig.\ref{Fig:slow-osc}.

Furthermore, if $\phi_1$ is as large as the Planck scale, there will be
 a stage of slow-roll inflation, as is always seen in conventional
 scenarios of chaotic inflation.
In this case, fast-roll (or locked) inflation starts after the end of
slow-roll
inflation and helps the precedent inflation by adding extra e-foldings
or reducing the vacuum energy and the reheat temperature.

\section{Models}
Assume that during inflationary period the potential is given by
\begin{equation}
V(\phi_1, \phi_2) = \frac{1}{2}m_1^2 \phi_1^2+\frac{1}{2}m_2^2 \phi_2^2.
\end{equation}
Introducing a complex field $\hat{\phi}\equiv \phi_1 + i \hat{A} \phi_2$,
where $\hat{A}$ is the ratio given by $\hat{A}=m_2/m_1$,
we can rewrite the potential as
\begin{equation}
V(\hat{\phi}) = \frac{1}{2}m_1^2 |\hat{\phi}|^2.
\end{equation}
Let me introduce a scalar field $\varphi$ that is given by
\begin{equation}
\hat{\phi} \equiv v e^{i\frac{\varphi}{v}},
\end{equation}
where $v$ is defined as $v=|\hat{\phi}|$.
Now the definition of the equipotential surface is
\begin{equation}
V = \frac{1}{2} m_1^2 v^2 = const.,
\end{equation}
and the field along the equipotential surface is $\varphi$.
We can calculate the potential for the field $\varphi$ that is defined
along the
equipotential surface, which is given by
\begin{equation}
V(\varphi)=\frac{1}{2}m_1^2 v^2 = const.
\end{equation}
It is easy to calculate the first derivative with respect to the field
$\varphi$, which is given by
\begin{equation}
\frac{\partial V(\varphi)}{\partial \varphi} =0,
\end{equation}
and the second derivative is given by
\begin{equation}
\frac{\partial^2 V(\varphi)}{\partial \varphi^2} =0,
\end{equation}
which shows that along these direction perturbations of 
the order of $H/2\pi$ will be generated,
since the second derivative is vanishing along the 
equipotential surface.

In this paper, we will consider cases in which $\phi_1$ is an actual
 inflaton that determines e-foldings, while $\phi_2$ is an additional
 field that is not relevant to the e-foldings.\footnote{To be more
 precise, $|\phi_2| \ll \phi_c/\sqrt{B}$ occurs before the  
end of inflation.}

If the fluctuations along the equipotential surface
 are sufficiently small compared with $\phi_i$,
one can decompose it along its tangent line.
Then, the fluctuations will be given by
\begin{equation}
\label{pert}
\left(\delta \phi_1, \delta \phi_2 \right)\simeq 
\frac{H_I}{2\pi \phi_{r,0}}
\left(A \times \phi_{2,0}, \phi_{1,0}\right),
\end{equation}
where $\phi_{i,0}$ denotes the value of $\phi_i$ at the time when 
fluctuations exit horizon.
In any case, the evolution of the fields $\phi_i$ is determined by 
Klein-Gordon equation (\ref{Klein-Gordon}).
Fluctuation of a light field is defined on any
spacetime slicing and is almost massless.
After horizon exit, the vacuum fluctuations have generated classical
perturbations, which vary slowly on the Hubble
timescale and are expected to be Gaussian.
These classical perturbations have the spectrum 
${\cal P}_{\delta \phi}^{1/2}=H_k/2\pi$,
where $H_k$ denotes the Hubble constant $H$ at horizon exit.
The evolution of $\zeta$ before horizon entry is described by the famous
$\delta N$ formalism\cite{delta-N,delta-N2,delta-N3}.
The time-dependent curvature perturbation smoothed on the scale $k$ is
\begin{equation}
\zeta(x,t)=\delta N(k,\phi_i(x), \rho(t)),
\end{equation}
where $N$ denotes the number of e-foldings from the epoch of horizon
exit to an epoch when the energy density has a value
$\rho$.\footnote{Note that in general $N$ depends on both $\phi_i$ and
$\dot{\phi}_i$, but if $t$ is during slow-roll inflation one can use the
slow roll approximation to eliminate the dependence on $\dot{\phi}_i$.
In our case, however, we cannot simply ignore the dependence because we
are not thinking about slow-roll inflation.
In the following calculations we will omit the $\dot{\phi}_i$-dependence
if it is obvious.
Be sure that we are not omitting the $\dot{\phi}_i$-dependence 
because of the slow-roll conditions, but considering the explicit form
of $N$ to see the $\dot{\phi}_i$-dependence.}
Using the above idea of the $\delta N$ formalism, Lyth made a concrete 
example for generating curvature perturbation at the end of
inflation\cite{delta-N2}.
To illustrate the idea, let us consider a model in which there is a
unique inflationary trajectory.
Then, $\phi_i$ becomes a unique function of the proper time $\tau$ up to a
shift in the origin.
As a result, all functions of $\phi_i$ will also be unique, including the
energy density $\rho=V+\frac{d^2\phi_i}{d^2 \tau}$.
In this case, at each position of the Universe inflation ends at
$\phi_I=\phi_{Ie}$ that is a constant and is independent of position.
The novel possibility discussed in ref.\cite{delta-N2} is that 
$\phi_{Ie}$ depends on some other field $\phi_{add}$ whose potential is
practically flat.
Then, $\phi_e$ depends on spatial position because of the perturbation
$\delta \phi(x)_{add}$.
This changes the e-foldings from a spacetime slice of uniform density
just before the end of inflation to a spacetime slice of uniform density
just after the end of inflation.
In this paper, the above change of the e-folds is denoted specifically
by $\delta N_{\Delta}\equiv \zeta_\Delta(x)$. 
This quantity $\zeta_\Delta$ is the contribution to the curvature
perturbation generated at the end of inflation.

We must be careful about the evolution of fluctuations after horizon
exit.
Although $\delta N_{\Delta}$ is produced at the end of inflation,
fluctuations $\delta \phi_{add}$ are produced much before the end of
inflation and have exited horizon during inflation. 
Here we would make a modest assumption that the value 
$\delta \phi_e(t)/\phi_e(t)$ is almost a constant during the
era\cite{EU-book, curvaton_3}. 
Thus, the fluctuation of $\phi_{add}$ at the end of inflation is
given by
\begin{equation}
\delta \phi_{add}(t_{e}) \simeq \frac{H_I(t_k)}{2\pi} 
\left(\frac{\phi_{add}(t_e)}{\phi_{add}(t_k)}\right),
\end{equation}
where $t_k$ and $t_e$ are the time when fluctuation exits
horizon and the time when inflation ends, respectively.
If the field $\phi_{add}$ is much lighter than $\phi_I$, $\delta
\phi_{add}(t_{e}) \simeq \delta \phi_{add}(t_{k})$ is a conceivable
approximation. 

\subsection{Fast$\times$flat}
\begin{figure}[ht]
 \begin{center}
\begin{picture}(400,200)(0,0)
\resizebox{15cm}{!}{\includegraphics{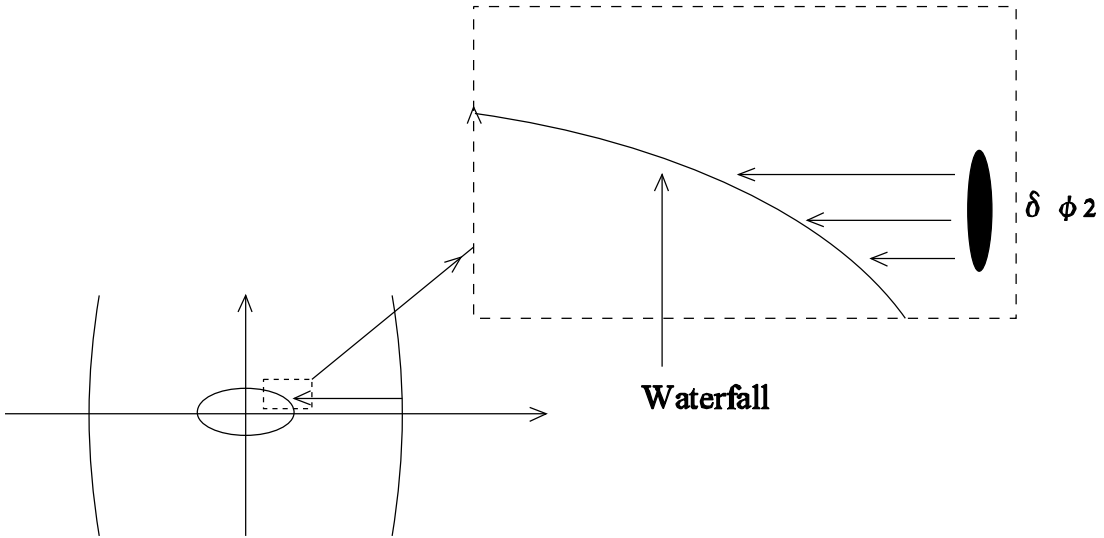}} 
\end{picture}
\caption{$\delta \phi_2$ leads to the delay of the waterfall stage.
This phenomenon looks similar to the usual back-and-force perturbation of
the single-field model, although inflaton field $\phi_1$ does not
roll slowly.
Of course, unlike the usual single-inflaton models, there is no reason
  that one can naively believe in (almost) scale-invariant and 
(almost) Gaussian perturbations.
It is lucky that we could find scale-invariant Gaussian perturbation in 
a certain parameter region of this model.}
\label{Fig:waterfall}
 \end{center}
\end{figure}

The easiest example would be that $\phi_1$ is fast-roll inflaton,
while $\phi_2$ is flat direction that does not roll during inflation.
To be more precise, $\phi_1$ and $\phi_2$ corresponds to $\phi_I$ and
$\phi_{add}$ in the above argument, respectively.  
In our scenario of elliptic inflation, this situation corresponds to 
$A\ll 1$ and $\phi_{2e} \ll \phi_c/\sqrt{B}$, 
where the equipotential surface becomes a squeezed ellipsoid.
Initially, the equipotential surface is given by
\begin{equation}
\phi^2_0 \equiv \phi_{1,0}^2 +A \phi_{2,0}^2.
\end{equation}
As one can see from Fig.\ref{Fig:waterfall}, the fluctuation along
equipotential surface is almost $\delta\phi_2\simeq H_I/2\pi$ due to the
above conditions $A\ll 1$ and $\phi_{2e} \ll \phi_c/\sqrt{B}$.

Because of the assumption $A\ll 1$ for the potential (\ref{pot-full}),
there is a large hierarchy between $m_1$ and $m_2$.
Thus, we can neglect the roll of the secondary field $\phi_2$ during
inflation.
Then, we can calculate $\zeta_\Delta(x)$ using the method advocated in
ref.\cite{delta-N2}, which takes the form of 
\begin{equation}
\label{toy1}
\zeta_\Delta(x) = \frac{\partial N_{\Delta}}{\partial \phi_1}
 \frac{\partial \phi_1}{\partial \phi_2} \delta \phi_2
+ \frac{1}{2}\left\{
	      2 \frac{\partial^2 N_{\Delta}}{\partial \phi_1^2}
	      \left(
\frac{\partial \phi_1}{\partial \phi_2}
\right)^2
	      + \frac{\partial N_{\Delta}}{\partial \phi_1} 
	      \frac{\partial^2 \phi_1}{\partial \phi_2^2}
\right\}(\delta \phi_2)^2.
\end{equation}
We do not assume slow-roll for the inflaton field $\phi_1$.
It is amazing that this mechanism for generating the curvature
perturbation still works to generate scale-invariant and Gaussian
perturbation, even in the case where $\phi_1$ is not a
slow-roll inflaton field.\footnote{This possibility is already suggested
in ref.\cite{delta-N2}.} 
As we will show later in this paper, this model presents a peculiar
limit of our more generalized scenario.
If $\phi_1$ is a fast-roll inflaton field, it is easy to obtain
\begin{eqnarray}
\left.\frac{\partial N_{\Delta}}{\partial \phi_1}\right|_e 
&=& -\frac{1}{F_1 \phi_{1e}} 
\nonumber\\
\left.\frac{\partial^2 N_{\Delta}}{\partial \phi_1^2}\right|_e &=& \frac{1}{F_1 \phi_{1e}^2},
\end{eqnarray}
and
\begin{eqnarray}
\left.\frac{\partial \phi_1}{\partial \phi_2}\right|_e
&=&-B \frac{\phi_{2e}}{\phi_{1e}}\nonumber\\
\left.\frac{\partial^2 \phi_{1}}{\partial \phi_{2}^2}\right|_e
&=&-\frac{B}{\phi_{1e}}\left(\frac{\phi_{2e}^2}{\phi_{1e}^2}+1\right).
\end{eqnarray}
As a result, $\zeta_\Delta(x)$ is given by
\begin{equation}
\label{zeta1}
\zeta_\Delta(x) = \frac{B H_I}{2\pi F_1 \phi_{1e}^2}
\left[ \phi_{2e}
+\frac{H_I}{4\pi} \left(\frac{2B \phi_{2e}^2}{\phi_{1e}^2}
+\frac{\phi_{2e}^2}{\phi_{1e}^2}+1 
\right)
\right],
\end{equation}
where we used $\phi_{2,0}\simeq\phi_{2e}$ to obtain $\delta
\phi_{2e}$. 
From the above equation, one can easily understand that non-Gaussianity 
parameter becomes large if $\phi_{2e} \ll H_I$.
As is seen from the above result (\ref{zeta1}), the curvature
perturbation does not depend on the value of $\phi_{i0}$.
Thus, at least in the limiting case where $A\ll 1$ is assumed, 
$\zeta_\Delta(x)$ can become scale-invariant in a sense that it does not
depend on the time when fluctuations exit horizon. 
Application of the above scenario is discussed in Sect.3.5 as a solution
to the eta-problem in brane inflationary models.
On the other hand, if $A$ is not so small, $\zeta_\Delta(x)$ may depend on
$\phi_{i0}$, as we will discuss below.

\subsection{Fast$\times$fast}
\begin{figure}[ht]
 \begin{center}
\begin{picture}(400,220)(0,0)
\resizebox{11cm}{!}{\includegraphics{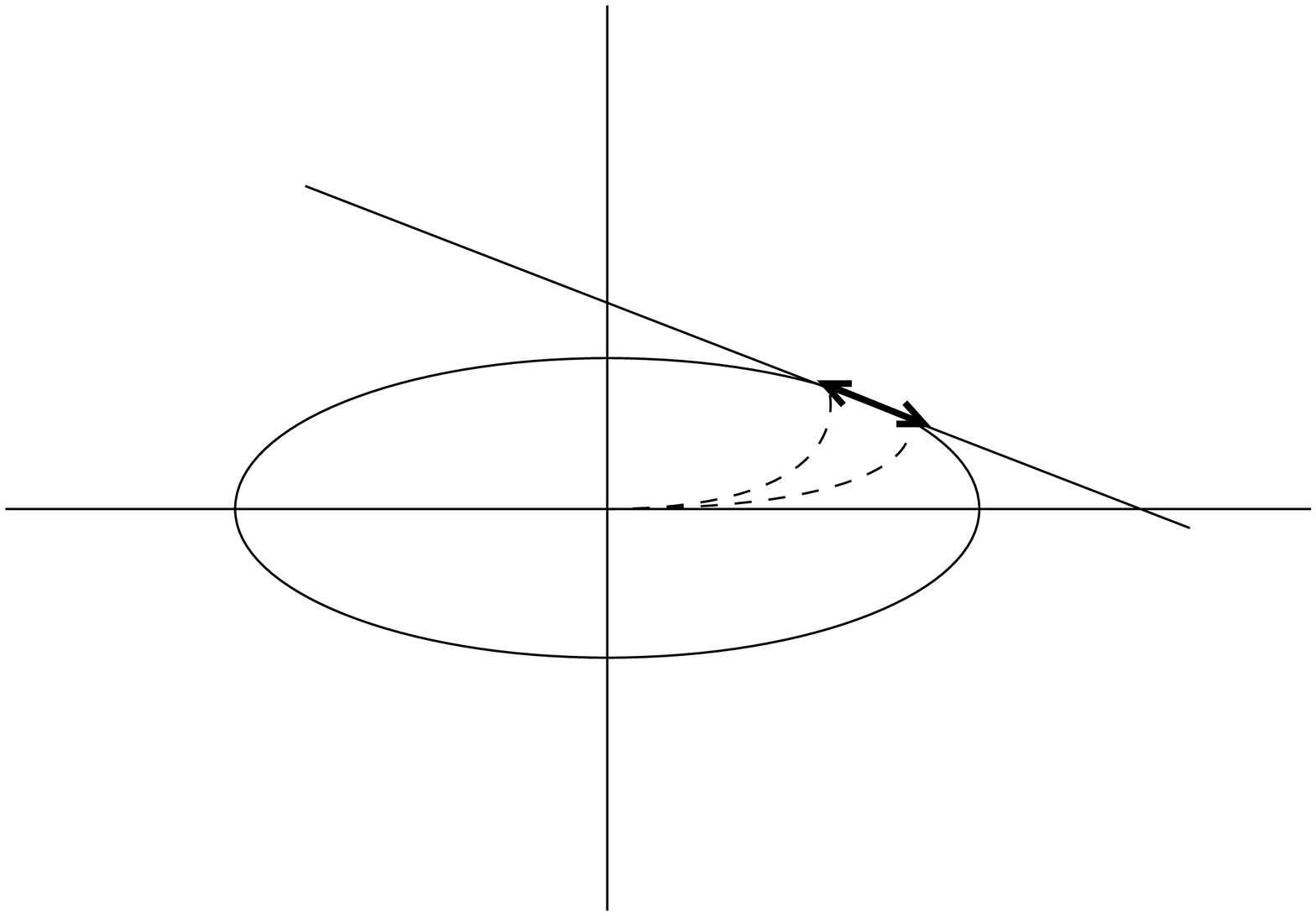}} 
\end{picture}
\caption{Fluctuations can appear along the equipotential surface even if
  the potential is steep for both $\phi_1$ and $\phi_2$.}
\label{Fig:elliptic}
 \end{center}
\end{figure}

What happens if we remove  the above condition $A\ll 1$ ?
Assume that the fields are initially at $(\phi_{1,0}, \phi_{2,0})$
and also that the order of magnitude of fluctuations are much smaller
than their initial 
values.
Then, one can decompose fluctuations along the tangent line of the
equipotential surface.
The fluctuations along the tangent line would be given
by eq.(\ref{pert}).
Although both $\phi_i(t)$ and $\delta \phi_i(t)$ will evolve during
inflation, it is conceivable to assume that the ratio $\delta
\phi_i/\phi_i$ is a constant during the inflationary
period\cite{curvaton_3}.   
Following the $\delta N$ formalism\cite{delta-N2}, it is easy to
calculate 
\begin{eqnarray}
\delta N&=&
\frac{\partial N_{\Delta}}{\partial \phi_1} \delta \phi_1+
\frac{\partial N_{\Delta}}{\partial \phi_1}
 \frac{\partial \phi_1}{\partial \phi_2} \delta \phi_2\nonumber\\
&&
+\frac{1}{2} \frac{\partial^2 N_{\Delta}}{\partial \phi_1^2} \left(
\delta \phi_1\right)^2
+ \frac{1}{2}\left\{
	      2 \frac{\partial^2 N_{\Delta}}{\partial \phi_1^2}
	      \left(
\frac{\partial \phi_1}{\partial \phi_2}
\right)^2
	      + \frac{\partial N_{\Delta}}{\partial \phi_1} 
	      \frac{\partial^2 \phi_1}{\partial \phi_2^2}
\right\}(\delta \phi_2)^2.
\end{eqnarray}
It would be important to note here that the initial fluctuations $\delta
\phi_i$ are not vanishing even if their masses are heavy.
See Fig.\ref{Fig:elliptic} for a perspective sketch of what happens in our
scenario.
Amazingly, our mechanism for generating the curvature perturbation
works even if both fields are massive.
We can use previous results
\begin{eqnarray}
\left.\frac{\partial N_{\Delta}}{\partial \phi_1}\right|_e 
&=& -\frac{1}{F_1 \phi_{1e}} 
\nonumber\\
\left.\frac{\partial^2 N_{\Delta}}{\partial \phi_1^2}\right|_e 
&=& \frac{1}{F_1 \phi_{1e}^2},
\end{eqnarray}
and
\begin{eqnarray}
\left.\frac{\partial \phi_1}{\partial \phi_2}\right|_e
&=&-B \frac{\phi_{2e}}{\phi_{1e}}\nonumber\\
\left.\frac{\partial^2 \phi_{1}}{\partial \phi_{2}^2}\right|_e
&=&-\frac{B}{\phi_{1e}}\left(\frac{\phi_{2e}^2}{\phi_{1e}^2}+1\right)
\end{eqnarray}
to calculate the curvature perturbation.
It gives 
\begin{eqnarray}
\label{toy2eq}
\zeta_\Delta(x) &=&-\frac{1}{F_1 \phi_{1e}}\left(\frac{H_I u_1}{2\pi}
\frac{\phi_{1e}}{\phi_{1,0}}\right)
+\frac{1}{F_1 \phi_{1e}}\left(B\frac{\phi_{2e}}{\phi_{1e}}\right)
\left(\frac{H_I u_2}{2\pi}
\frac{\phi_{2e}}{\phi_{2,0}}\right)
\nonumber\\
&&\frac{1}{2F_1 \phi_{1e}^2}\left(\frac{H_I u_1}{2\pi}
\frac{\phi_{1e}}{\phi_{1,0}}\right)^2\nonumber\\
&&\frac{1}{F_1 \phi_{1e}^2}\left(\frac{B\phi_{2e}}{\phi_{1e}}\right)^2
\left(\frac{H_I u_2}{2\pi}
\frac{\phi_{2e}}{\phi_{2,0}}\right)^2
-\frac{B}{2F_1 \phi_{1e}^2}
\left(\frac{\phi_{2e}^2}{\phi_{1e}^2}+1\right)
\left(\frac{H_I u_2}{2\pi}
\frac{\phi_{2e}}{\phi_{2,0}}\right)^2,
\end{eqnarray}
where we have defined $u_1\equiv \frac{A \times
\phi_{2,0}}{\phi_{r0}}$  
and $u_2 \equiv \frac{\phi_{1,0}}{\phi_{r0}}$.
The first line of eq.(\ref{toy2eq}) vanishes when $A=B=1$, as is
expected, while the second-order terms do not.
Remember that we have truncated higher order terms when we obtained
the perturbation eq.(\ref{pert}).
Now it is obvious that we should have included higher terms in
eq.(\ref{pert}) to obtain exact result for second order terms in
eq.(\ref{toy2eq}).
On the other hand, the result (\ref{zeta1}) is a good approximation,
since we have assumed $A \ll 1$ in Sect.3.1.
At this time, we are not interested in the second order terms in the
above result.
Obviously, the result (\ref{toy2eq}) depends on the time when
fluctuations exit horizon, since in the above equation there are terms 
that depend on $\phi_{i0}$.
Note that $\phi_{i0}$ parameterizes the time when fluctuations exit
horizon. 
According to eq.(\ref{toy2eq}),  perturbations produced at later
stages of elliptic 
inflation have much larger amplitudes in its spectrum, thus it has a
blue spectrum.
This is a problem for the people who want to generate the curvature
perturbation only from elliptic inflation, while it seems a good idea
to obtain a tilted spectrum, at least for the people who want to embed
this idea into a model where the scale-invariant curvature perturbation
is generated by another mechanism.
For example, one may consider a model where slow-roll inflation occurs
before elliptic inflation.
We will discuss about this issue again in Sect.3.4.
Of course, it may be a good idea to obtain a tilted spectrum 
embedding this model into a scenario with the
curvatons\cite{Dimopoulos-Fast-Osc}. 
In this case, one may assume that the large-scale perturbation is
generated by the curvatons, while the perturbation generated by 
 elliptic inflation becomes large in its short-scale side, thus
 generating a tilted spectrum. 

\subsection{Fast$\times$locked}
\begin{figure}[h]
 \begin{center}
\begin{picture}(400,220)(0,0)
\resizebox{15cm}{!}{\includegraphics{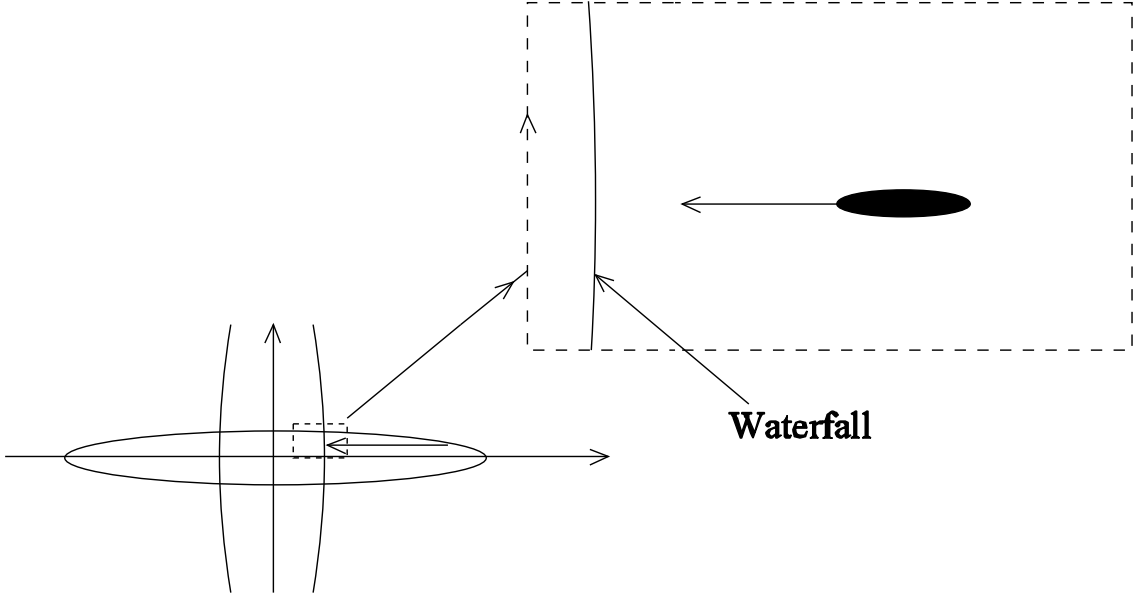}} 
\end{picture}
\caption{The component of the fluctuation is almost $\delta \phi_1\simeq \frac{H_I}{2\pi}$ because we took a limit $A \gg 1$
(i.e. the additional field $\phi_2$ is very heavy.).
The fluctuation $\delta \phi_1$ leads to the delay of the waterfall stage,
  which looks 
precisely the same as the usual back-and-force perturbation of
slow-roll inflation.
In this case, however, one cannot expect scale invariance because 
$\phi_1$ does not roll slowly.}
\label{Fig:toy2}
 \end{center}
\end{figure}

It would be useful to consider the opposite limit of $A\ll 1$,
where the additional field $\phi_2$ is very massive.
Of course, we do know it is quite normal to think that an additional
field cannot play any role in generating the curvature perturbation if
it is heavier 
than the inflaton field.
However, this anticipation is not always
true in our scenario.
The idea we will consider here is schematically sketched in
Fig\ref{Fig:toy2}, where  
$B \ll 1$ and $ A\gg 1$ is assumed.
In this case, we can take $u_2 \simeq 0$ and $u_1
\simeq 1$, and can neglect all terms that are proportional to $u_2$.
Then, from eq.(\ref{toy2eq}) we obtain
\begin{eqnarray}
\zeta_\Delta(x) &\simeq&\frac{H_I }{2\pi F_1 \phi_{1,0}}
-\frac{(H_I )^2}{8\pi^2 F_1 \phi_{1,0}^2},
\end{eqnarray}
where the higher terms that are hidden in eq.(\ref{pert}) can be
neglected, since 
we are considering an extreme case with $A \gg 1$.
This result is quite interesting.
We have obtained a blue spectrum in an inflationary scenario
without slow-roll, by just adding a heavy field $\phi_2$ that couples
very weakly to $\sigma$.
A plausible situation is that the secondary field $\phi_2$ 
oscillates rapidly during inflation, while $\phi_1$ play the role of 
the conventional fast-roll inflaton field.\footnote{See also
Fig.\ref{Fig:osc_fast}.} 
Then, fluctuations will be generated along the direction that is almost
perpendicular to the oscillation.
This direction is almost massless, like fluctuation of a
falling raindrop.
At the end of inflationary expansion, $\delta N_\Delta$ is produced by
the fluctuation.
Although the expectation value of the heavy field $\phi_2$ may
already be vanishing at the end of inflation and may not appear
explicitly in the effective action at this time, one cannot disregard 
the fluctuations induced by this field, which are generated and 
exited horizon during earlier period of inflation.

In some cases, one may consider fast-roll (or locked) inflation as a
source for e-folds, while expecting that the curvature perturbation is
produced by the curvatons. 
Then, one will disregard the curvature perturbation that may
be generated during the period of fast-roll (or locked) inflation.
This scenario seems plausible, however the situation is not so simple as
it has been anticipated before.
To illustrate our idea, let us include a heavy field $\phi_2$ in
addition to the fast-roll (or locked) inflaton field $\phi_1$.
As we have seen above, a heavy field may oscillate at the beginning of
inflation, producing a blue spectrum.
In the large-scale side, the spectrum may be dominated by the
curvature perturbation generated by the curvatons as expected, however
its short-scale side may be dominated by the curvature perturbation
generated at the end of fast-roll inflation.
This scenario gives a tilted spectrum.

\subsection{Built-in scenarios}
As we have discussed above, it is not easy to obtain scale-invariant
spectrum only from elliptic inflation.
However, it would be interesting if elliptic inflation starts after
a short period of slow-roll inflation.
To illustrate our idea, let us assume that the (scale-invariant)
perturbation is produced during a precedent stage of slow-roll inflation
that generates minimum amount of e-foldings. 
Then, one obtains a tilted spectrum, which is due to the succeeding
stage of elliptic inflation.

We think it is quite natural to expect the occurrence of fast-roll (or
locked) inflation after the period of conventional slow-roll
inflation.\footnote{For example, brane inflation is supposed to end when
the inflaton mass exceeds the Hubble constant, which may be
accompanied by fast (or locked) inflationary phase.}
If so, the spectrum should have unusual tilt in its short-scale side.
Today, the smallest scale accessible to large-scale structure
observations is about 1Mpc.
This scale is equal to the Hubble radius about nine e-foldings after
our present Hubble scale $a_0 H_0$ exit horizon.
Therefore, we need at least nine e-foldings of slow-roll expansion that
generates scale-invariant spectrum, if it is to explain the observed
large-scale structure.

\begin{figure}[h]
 \begin{center}
\begin{picture}(400,300)(0,0)
\resizebox{15cm}{!}{\includegraphics{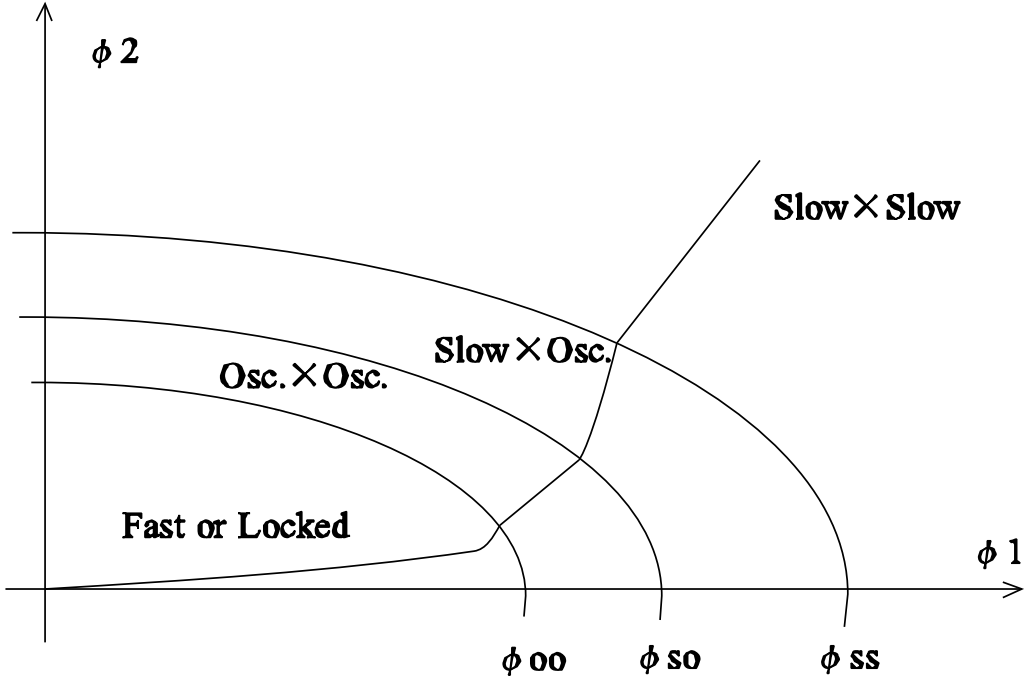}} 
\end{picture}
\caption{A schematic picture of the ``built-in'' scenario.
Here we set $A>1$.
Slow-roll phase at a distance is important in producing the
large-scale spectrum, if there is no curvatons.} 
\label{Fig:built-in}
 \end{center}
\end{figure}

Then, it would be interesting if one can find signatures of the
succeeding stage of (secondary) inflation from the intermediate-scale
cosmological observations.
To examine this idea, let us consider a potential
\begin{equation}
\label{pot2}
V_{I} \simeq V_0 + \frac{1}{2}m \phi_r^2.
\end{equation}
In the above calculations, we simply assumed that $V_0$ is much
larger than mass terms $\frac{1}{2}m^2_i \phi_i^2$ and dominates the
vacuum density during inflation.
However, the above assumption is violated if $m$ and $\phi_r$ are
large, and if the second term dominates the potential.
In the case where the mass term dominates the vacuum density, the Hubble
constant is 
given by $H_I \simeq m \phi_r /M_p$.
This happens in the region where $\phi_r\ge\phi_{oo}\equiv \sqrt{V_0}/m$.
We define $\phi_{so}$ and $\phi_{ss}$  so that 
$H_I > m_1$ and $H_I > m_2$ are satisfied when
$\phi_r > \phi_{so} \simeq M_p$ and $\phi_r > \phi_{ss} \simeq
\sqrt{A}M_p$, respectively.
Elliptic inflation starts after slow-roll inflation, at the time when
$\phi_r < \phi_{oo}$.  
$\phi_1$ (and not $\phi_2$) becomes slow-roll if
$\phi_{so} \le\phi_r\le\phi_{ss}$, and both $\phi_1$ and $\phi_2$
rolls slowly at a distance $\phi_r > \phi_{ss}$.
See also Fig.\ref{Fig:built-in}

\subsection{Brane inflation}
Although brane inflation is in a sense 
a variant of hybrid inflation, it may look
different from the original.
For example, in brane inflationary scenarios waterfall is
triggered when the branes come closer than a critical distance.
The mass of the waterfall field and the critical distance is determined
by string dynamics on the branes, while the mass of inflaton (i.e. the
mass of the position moduli) is determined by the global and local 
structure of the extra dimensions.
Although brane inflation is a promising idea that explains how inflation
is realized in brane cosmology,
 it is known that obtaining successful flat direction is quite
difficult\cite{KKLT-eta}.
From the above viewpoints, it is important to examine if the curvature
perturbation can be produced without slow-roll, at the same time
without curvatons.
It would be useful to note here that the
``brane distance'' is given by $r^2=\sum_i x_i^2$, where $x_i$ are the
brane distances in the directions perpendicular to the inflating branes.
In common case, there is no exact symmetry that protects all the
masses of the position moduli.
Thus, the equipotential surface of such potential is elliptic.
As we have discussed above, one tends to think that the secondary field
(or direction in extra dimensions perpendicular to the inflaton) plays
no role in generating the curvature perturbation, if it
is heavier than the primary inflaton field.
However, this is not the case.
One cannot disregard other inflaton candidates that are much
heavier than the actual inflaton.
Although brane inflation is supposed to end when the effective mass
becomes as large as the Hubble constant, there may exist another period of
inflationary expansion (fast or locked inflation) just after slow-roll
inflation.
In this sense, what we have discussed above in this paper 
provides us with a new inflationary scenario that makes it possible to
generate a tilted spectrum in brane inflationary models.
All flat directions can be lifted at the time of elliptic inflation.
It does not disturb our mechanism for generating the curvature
perturbation. 

On the other hand, our mechanism is useful in solving the serious
eta-problem in brane inflationary models.
To illustrate our idea, let us consider a concrete example.

\subsubsection{Fast$\times$flat inflation in a brane inflationary model}

\begin{figure}[h]
 \begin{center}
\begin{picture}(400,330)(0,0)
\resizebox{13cm}{!}{\includegraphics{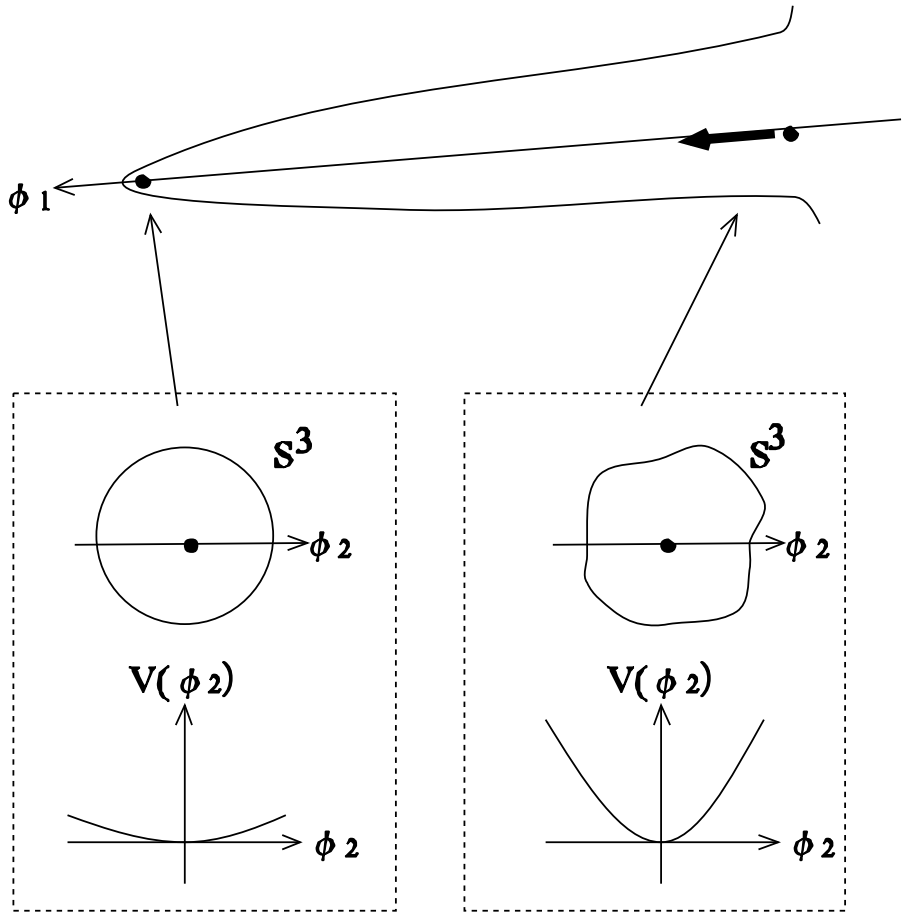}} 
\end{picture}
\caption{The deformation of the KS throat at the UV may explicitly
  breaks the $SO(4)$ symmetry, generating masses for the corresponding
  moduli. On the other hand, the deformation is milder at the tip of the
  throat, keeping the $SO(4)$ symmetry at the tip.
  Be sure that there is no reason that one must use the fluctuation of
  the moving brane in calculating the generation of the curvature
  perturbation.}
\label{Fig:throat}
 \end{center}
\end{figure}

The setting for our analysis is the dS background of
KKLT\cite{KKLT-original}.
The stabilization of the K\"ahler modulus leads to a vacuum with a
negative cosmological constant, however the vacuum can be lifted by
adding an $\overline{D3}$-brane that sits at the tip of the throat.
An important consequence of the introduction of an $\overline{D3}$-brane
to the warped background is the appearance of new moduli that
correspond to the position of the  $\overline{D3}$-brane  on the compact
space.
The potential for these moduli in the KKLT background is rather curious.
One can see from ref.\cite{KKLT-eta,KKLT-original} that the
$\overline{D3}$-brane is not free to move in the throat direction since
they have a potential proportional to the warp factor.
This potential stabilizes the $\overline{D3}$-brane at the tip of the
throat.
However, at the tip of the throat, the $\overline{D3}$-brane can still
move on the $S^3$.
In the exact KS solution, $S^3$ is exact at the tip of the throat and
the symmetry of $S^3$ is broken by placing the $\overline{D3}$-brane as
$SO(4)\rightarrow SO(3)$, which gives rise to three massless moduli.
The moduli correspond to the three coordinates of the position of the
$\overline{D3}$-brane on the $S^3$.
In generic background, there would be corrections from the UV-boundary,
the Calabi-Yau manifold, since the background deviates from the original
KS solution away from the tip.
These corrections may explicitly break the $SO(4)$ symmetry, thus
generate masses for the moduli.
In ref.\cite{mass-KS}, one can find calculations for the masses that
arise from the UV-corrections.
Although the deformation of the theory at the UV-side
generates massses for the moduli, the highest contribution is
exponentially smaller than the typical mass scale at the tip.
Therefore, in a brane inflationary scenario where the
$\overline{D3}$-brane is fixed at the tip of the throat while the moving
brane comes from the root, the fluctuation of the $\overline{D3}$-brane
in the direction of $S^3$ becomes important.
In our ``Fast$\times$flat scenario'', the throat direction and the $S^3$
moduli correspond to the fields $\phi_1$ and $\phi_2$ ,
respectively.\footnote{See also Fig.\ref{Fig:throat}.}
Thus, the fluctuation of $\overline{D3}$-brane at the tip can generate the
curvature perturbation even if all the scalar fields parameterizing
the position of the moving brane is lifted to have large masses.
Although it will be rather difficult to explain total 
e-foldings only from single-shot inflation, one may add secondary
inflation that may take place after fast-roll inflation.
Thus, we can solve the eta-problem without using the curvatons.
We think our idea is important in constructing inflationary models
in KKLT and other scenarios.

\section{Conclusions}
The most important point is that an equipotential surface
appears whenever there is additional inflaton that couples to waterfall
field.  
Fluctuation appears along the equipotential surface and exits
 horizon during inflation. 
The equipotential surface is an ellipsoid, along which
fields can fluctuate despite the large mass of the fields.
Then, it induces fluctuation of the total number
 of e-folds at the end of inflation.
The end line of hybrid inflation may also be ellipsoid,
 which becomes another source for the curvature perturbations.
The spectrum of the curvature perturbation is determined by the masses
 and the couplings of the fields.

For brane inflationary models, our mechanism provides us with a new
inflationary scenario that circumvent the eta-problem.
It is possible to generate the curvature perturbation even in a model
where all flat directions are lifted at the UV side of the KS throat.
It is also notable that our mechanism can be used to obtain a tilted
spectrum.
The spectrum is determined by the masses of the position moduli that
parameterize the position of the inflationary branes. 
The mechanism may provide us with an important clue as to
how one can obtain information about the structure of extra dimensions
from cosmological observations.

\section{Acknowledgment}
We wish to thank K.Shima for encouragement, and our colleagues in
Tokyo University for their kind hospitality.

\end{document}